\documentstyle[aps,preprint]{revtex}
\begin{document}
\title{Comment on `Observation of Bose-Einstein Condensation of Molecules',
cond-mat/0311617}
\author{M. Wouters\thanks{%
M.W. and J.T. are supported financially by the FWO-Vlaanderen.}, J. Tempere$%
^{\ast }$, J.T. Devreese}
\address{TFVS, Universiteit Antwerpen, Universiteitsplein 1, B2610 Wilrijk,
Belgium. }
\date{December 9, 2003.}
\maketitle

\bigskip

\tighten Zwierlein {\em et al.} report in their recent preprint \cite%
{Zwierleincondmat0311}\ a value for the scattering length of weakly bound
molecules that are formed out of two fermionic $^{6}$Li atoms in different
internal states. They observe a Thomas-Fermi density profile typical for a
condensate of molecules, but there seems to be a serious discrepancy between
the measured value of the measured molecular scattering length of $a_{\text{%
mol}}^{\text{exp}}=8$ nm \cite{Zwierleincondmat0311,Ketterlecomm} at a
magnetic field of 770 G and the predicted one \cite{Petrovcondmat0309} of $%
a_{\text{mol}}^{\text{theory}}=0.6$ $a_{\text{atom}}=120$ nm. In this
comment, we argue that this might be an indication that they have produced
in fact a new many-body state.

\bigskip

As a first guide to explain this discrepancy, we want to point out that the
Bose-Einstein condensed molecular gas is not dilute, in the sense that $%
n^{1/3}R$ is not very small ($R$ is the radius of the molecule and $n$ the
density). The reported density is $5\times 10^{13}$cm$^{-3}$ in the center
of the trap and the measured radius of the molecule $R\geq 100$ nm,
consistent with the predicted value for the atomic scattering length $a_{%
\text{atom}}=200$ nm. This gives us $n^{1/3}R=0.368$. The supposition that
the molecular gas behaves as if the molecules were pointlike bosons is then
not necessarily fulfilled and in some sense this experiment is already in
the cross-over regime between BEC and BCS and not anymore in the true BEC
limit. This might explain the discrepancy between the theory and the
experiment, because the theory of Petrov, Salomon and Schlyapnikov. \cite%
{Petrovcondmat0309} only deals with the true BEC limit, for which it
predicts $a_{\text{mol}}^{\text{BEC}}=0.6$ $a_{\text{atom}}$.

Apart from the result of Ref. \cite{Petrovcondmat0309} for molecules, there
is a second class of papers that discusses the cross-over between BEC and
BCS starting from the BCS-side \cite{Legget1980,EngelbrechtPRB55,PieriPRL91}%
\ in stead. In these papers, a BCS ansatz is made for the wave function of
the many body system. This BCS-ansatz seems to give meaningful results both
in the weak-coupling (BCS) and strong-coupling (BEC) limit, predicting for
the molecule-molecule scattering length the value of $a_{\text{mol}}^{\text{%
BCS}}=2a_{\text{atom}}$. This would result in a molecular scatting length of
400 nm in this case, even further away from what is observed experimentally.

\bigskip

Insight can be gained by a closer look at the way in which the scattering
length was determined experimentally, namely through a measurement of the
mean-field energy \cite{Zwierleincondmat0311}. To compare the experimental
result with that of the two theories ($a_{\text{mol}}^{\text{BEC}}$\ from %
\cite{Petrovcondmat0309}, vs. $a_{\text{mol}}^{\text{BCS}}=2a_{\text{atom}}$%
\ from \cite{Legget1980,EngelbrechtPRB55,PieriPRL91}), we have to start from
the mean-field energy predicted in these theories.

As stated, in order to obtain the scattering length from the experiment, the
mean-field energy is measured \cite{Zwierleincondmat0311}. On the one hand,
for a weakly interacting bose-gas, this mean-field energy equals 
\begin{equation}
\mu _{\text{mol}}^{\text{BEC}}\left[ n\right] =\frac{4\pi \hbar ^{2}a_{\text{%
mol}}n}{m_{\text{mol}}}.  \label{muboson}
\end{equation}%
This is the chemical potential of the homogeneous Bose-gas with density $n$.
In the Thomas-Fermi approximation of describing the condensate, this is the
only quantity that has to be known, since one assumes that 
\begin{equation}
\mu _{\text{mol}}^{\text{BEC}}\left[ n\left( {\bf r}\right) \right] +V_{%
\text{ext}}\left( {\bf r}\right) =\lambda ,  \label{Thomas-Fermi}
\end{equation}%
with $V_{ext}$ the external trapping potential an $\lambda $ the chemical
potential of the trapped Bose gas. This allows us the calculate all the
properties, such as the density profile $n\left( {\bf r}\right) $ and the
total potential energy. The only assumption on which (\ref{Thomas-Fermi}) is
based is that $\mu \left[ n\left( {\bf r}\right) \right] \gg \hbar \bar{%
\omega},$ where $\bar{\omega}$ denotes the geometric mean of the three
trapping frequencies. The reported value of the mean field energy is 100 nK
and $\bar{\omega}=6.18$ nK, so that the Thomas-Fermi expression (\ref%
{Thomas-Fermi}) is valid.

On the other hand, we can also calculate the chemical potential starting
from a BCS-ansatz. Analytical results can be obtained both in the
strong-coupling{\bf \ }(BEC) and weak-coupling (BCS) limits. In the
strong-coupling BEC-limit, it is \cite{EngelbrechtPRB55} 
\begin{equation}
\mu _{\text{mol}}^{\text{BCS}}=2\mu _{\text{atom}}=-E_{\text{b}}+\frac{4\pi
\hbar ^{2}a_{\text{atom}}n}{m_{\text{atom}}},
\end{equation}%
with $E_{\text{b}}$ the molecular binding energy. This is equal to (\ref%
{muboson}), up to a density-independent constant, when we set $a_{\text{mol}%
}=2a_{\text{atom}}$ and $m_{\text{mol}}=2m_{\text{atom}}$, which agrees with
the results from Refs. \cite{EngelbrechtPRB55,PieriPRL91}. Ref. \cite%
{EngelbrechtPRB55} shows that in the case of the present experiments the
analytical estimation deviates from the numerical result. When we do the
numerical calculation for $a_{\text{atom}}=200$ nm and $n=5\times 10^{13}$cm$%
^{-3},$ we get $\mu _{\text{mol}}^{\text{BCS}}=6.4$ $\mu $K, which converted
to a scattering length with (\ref{muboson}) gives $a_{\text{mol}}^{\text{BCS}%
}=253$ nm, which is also not very close to the measured value.

The measured mean-field energy of a homogeneous molecular gas could be
calculated in the BCS-model, using 
\begin{equation}
E=\int_{0}^{n}\mu \left[ n^{\prime }\right] dn^{\prime }=\frac{2\pi \hbar
^{2}a_{\text{mol}}n^{2}}{m_{\text{mol}}}.
\end{equation}%
Keeping in mind the fact that the BCS-result is variational, we see that we
can only estimate a lower bound for $a_{\text{mol}}$. Apparently the
experiments of Zwierlein {\em et al.}{\bf \ }show that this estimation is
quite far off and this indicates that the BCS-wave function is probably far
from the true state of the system.

Drawing a similar conclusion for the prediction of Petrov {\it et al.} \cite%
{Petrovcondmat0309} is not on the same footing, because it is not clear what
many-body wave-function should be used in this case of a dense molecular
system ($n^{1/3}R=0.368$). Anyway, also in this case, if we naively use the
value of the molecule-molecule scattering length and ignore the dense
character of the system, we would arrive at a total mean-field energy that
is far higher than the measured energy and again conclude that this is an
indication that the many-body state cannot be very similar to a `Bose-gas of
molecules'.

\bigskip

In our view, the experiment of Ketterle and co-workers does not necessarily
need to be interpreted as indicating that the true scattering amplitude
between two molecules becomes small (8 nm measured vs. 120 nm predicted).
Rather it can be interpreted as indicating that the energy of the achieved
state is much lower than both the (variational) energy for a BCS state and
the energy of a molecular condensate. In this sense, we suggest that the
experiment may point to a novel many-body state, that is neither the BCS nor
the molecular BEC state, and that still has to be identified.

\end{document}